\input harvmac
\noblackbox
\Title{HUTP-98/A082}{High-Energy Tests of Lorentz Invariance}

\bigskip
\centerline{Sidney Coleman and Sheldon L. Glashow}
\bigskip
\centerline{Lyman Laboratory of Physics}
\centerline{Harvard University}
\centerline{Cambridge,  MA 02138}
\vskip .3in

We develop a perturbative framework with which to discuss departures from
exact Lorentz invariance and explore their potentially observable
ramifications. Tiny non-invariant terms 
introduced into the standard model Lagrangian
are assumed to be renormalizable
(dimension $\le 4$), invariant under $SU(3)\otimes SU(2)\otimes U(1)$ gauge
transformations, and rotationally and translationally invariant in a
preferred frame. There are a total of 46 independent CPT-even perturbations
of this kind, all of which preserve
anomaly cancellation.  They define the energy-momentum eigenstates and 
their maximal attainable
velocities in the high-energy limit. The effects of these perturbations
increase rapidly with energy in the preferred frame, more rapidly than
those of CPT-odd perturbations. Our analysis of Lorentz-violating
kinematics reveals several striking new phenomena
that are relevant both to cosmic-ray
physics ({\it e.g.,} by undoing the GZK cutoff) and neutrino physics  ({\it
e.g.,} by generating novel types of neutrino oscillations).  
These may lead to new and
sensitive high-energy tests of special relativity.

\Date{18/1/99}

\newsec{Introduction}

Experimental tests of Lorentz invariance have become remarkably accurate.
To give a quantitative measure of this accuracy, one imagines adding  tiny
Lorentz-violating terms to a  conventional Lagrangian. Experiments can test
Lorentz invariance by setting upper bounds to the coefficients of these
terms.  One common choice~\ref\rPT{{\it E. g.,} M. P. Hagen and C. M. Will,
Phys. Today 40 (May 1987), 69.} is to alter the 
coefficient of the square of the magnetic field
in the Lagrangian of
quantum electrodynamics:
\eqn\eIi{\vec { B}^2\rightarrow
(1+\epsilon) \vec { B}^2.} 
Among other effects, this term causes the
velocity of light $c$, given by $c^2=1+\epsilon$, to differ from the
maximum attainable velocity of a material body, which remains equal to one.
(Shortly we shall
consider more general Lorentz-violating perturbations.)

The perturbation \eIi~breaks Lorentz invariance. It is
translationally and rotationally  invariant in the frame
in which we are working (``the preferred frame'') but not in any other
frame. 
If the preferred frame is the
one in which the cosmic microwave background
is isotropic, tiny and calculable anisotropies should
appear in laboratory experiments.  High-precision spectroscopic experiments
that fail to find such anisotropies \ref\rLam{S. K. Lamoreaux, J. P.
Jacobs, B. R. Heckel, F. J. Raab, and E. N. Fortson, Phys. Rev. Lett. 57
(1986) 3125.} \ref\rberg{
 C.J. Berglund, L.R. Hunter, D. Krause, Jr., E.O. Prigge, and M.S.
Romfeldt,
Phys. Rev. Lett. 75
(1995) 1879.} set
the bound 
$\vert 1-c^2\vert=|\epsilon|<6\times 10^{-22}$.

In a note published last year~\ref\rCG{S. Coleman and S. L. Glashow, Phys.
Lett. B405 (1997) 249.} we  pointed out that a better bound could be
obtained 
from a very different sort of experiment if
$c<1$. In this case 
a charged particle traveling faster
than light rapidly radiates photons
until it is no
longer superluminal.
Thus no primary cosmic-ray proton can have energy
greater than $M/ \sqrt{1-c^2}=M|\epsilon|^{-1/2}$, where $M$ is the proton
mass.  Because primary protons with energies up to $10^{20}$ eV are seen, 
we set the bound
$1-c^2< 10^{-23}$, almost an order of
magnitude stronger than the atomic-physics bound.  High accuracy is
obtained from high energy rather than high precision.  Moreover our
bound requires no
assumption about our velocity in the preferred frame.

This effect, which we call
vacuum \v Cerenkov radiation, is absent below a characteristic
energy and turns on abruptly once that energy
is reached.  Such is not always
the case, as the following example shows.

Let $\Psi$ denote a set of $n$ complex scalar fields  assembled into a column
vector.  If we assume invariance under the $U(1)$ group $\Psi\rightarrow
e^{-i\lambda}\Psi$, the most general free Lagrangian is:
\eqn\eIii{{\cal L} =\partial_\mu\Psi^*Z\partial^\mu\Psi-\Psi^*M^2\Psi,}
where $Z$ and $M^2$ are positive Hermitian matrices.   We can always 
linearly  transform the fields
to make $Z$ the identity and $M^2$ diagonal,
thus obtaining the standard theory of $n$ decoupled free fields.
Consider adding to the Lagrangian the  Lorentz-violating term:
\eqn\eIiii{{\cal L}\rightarrow{\cal L} +
\partial_i\Psi \epsilon\partial^i\Psi,}
where $\epsilon$ is a Hermitian matrix. If $\epsilon$ does not commute
with $M^2$, there is no way to disentangle the fields.  The single-particle
energy-momentum eigenstates go from eigenstates of $M^2$ at low momenta to
eigenstates of $\epsilon$ at high momenta. In contrast to vacuum \v Cerenkov
radiation, this effect, which we call velocity mixing, turns on gradually.
Gradual effects allow one to obtain high accuracy by combining
moderately high energies with moderately high precision.

A more striking gradual effect appears if this system is
minimally coupled to electromagnetism.  In this case, a meson can decay to
a less massive meson plus a photon at a rate growing with the cube
of the energy.  Analogous terms in the standard model can
drive the otherwise-forbidden decay $\mu^+\rightarrow e^++\gamma$ 
and  the 0-0 transition $K^+\rightarrow \pi^++\gamma$.

This example shows that  what meant  by ``high energy'' 
in this context
depends very much on the details of the system under consideration.  For
simplicity, suppose $n=2$  and let $M^2={\rm diag}~(M_1^2,
M_2^2).$  The transition from eigenstates of $M^2$ to those of
$\epsilon$ occurs at energies 
$\sim \sqrt{(M_1^2-M_2^2)/\epsilon_{12}}$.  For the neutral kaon system,
this energy is many orders of magnitude less than the
characteristic energy of vacuum \v Cerenkov radiation (if the
dimensionless invariance-violating parameters in the two processes are 
comparable).

These are just illustrative examples.  In  \S 2 we
study all local Lorentz-invariance violating interactions that are
rotationally and translationally invariant in a preferred frame and
of renormalizable type ({\it i.e.,} having
mass dimension
$\leq 4$).\foot{The condition of renormalizability can be given the usual
justification:  If we assume that breaking of Lorentz invariance occurs at
some very small distance scale, only the renormalizable interactions survive
the renormalization-group running of the couplings to experimentally
accessible scales.}  
(Some of the results in \S 2.1 and \S 2.2 were first established by
Kosteleck\'y and Colladay~\ref\rkost{D. Colladay and
V.A. Kosteleck\'y, Phys. Rev. {D55} (1997) 6760, D58 (1998) 116002,
and earlier references therein.}.
We  derive them anew
here, both for completeness and because we wish to emphasize features
relevant to high-energy tests.)

Lorentz-violating perturbations can be divided into two classes, depending
on whether they are even or odd under CPT.   For a state with  energy $E$,
we show in \S 2.1 that the expectation values of the CPT-even interactions
grow like $E^2$ for large $E$,   while those of the CPT-odd interactions 
grow like $E$.  Because we are interested in effects of very weak
interactions made detectable by high energies, we limit ouselves  primarily
to a study of the the CPT-even  interactions.\foot{This argument would be
evaded if the CPT-even  couplings were on the order of the squares of the
CPT-odd ones, expressed as dimensionless ratios to an appropriate mass
scale. This would be consistent with both renormalization-group flow and
our energy-growth rule, and would lead to the dominance of the CPT-odd
interactions at moderately high energies.  We  mainly ignore this
possibility here.} (Both our examples are of this class.) In \S 2.2 we
construct the most general CPT-even interaction for the standard model
and show that it preserves anomaly  cancellation. In \S 2.3 we discuss
certain approximations  that we will use later on. As an illuminating
exercise, in \S 2.4  we work out the kinematics of $n$-body decays in the
special case in which off-diagonal matrix elements of the velocity-mixing
matrices may be neglected.  Novel phenomena arise.  For example,  a decay
can be kinematically allowed both at low and high energy, but forbidden for
an intermediate range of energies.

The last section applies our formalism to various possibly
observable  manifestations
of Lorentz violation. We discuss  phenomena involving charged leptons in \S
3.1, in particular the possible appearance of radiative muon decay at
high energies. We discuss  phenomena involving neutrinos in \S 3.2, where
we show how searches for neutrino oscillations at high energy
and long baseline can provide new and powerful  tests of special
relativity. We discuss hadronic manifestations of Lorentz-violation in \S
3.3, especially those relating to ultra-high energy cosmic rays. An
Appendix explains why, contrary to our earlier assertion~\rCG, the
observed absence of a velocity difference between right- and left-handed
photons  (which would violate CPT)  does not constrain the appearance of
CPT-violating effects elsewhere.

\newsec{Generalities}
Here we develop the formalism needed to determine the observable
consequences of Lorentz violation so as to obtain precise high-energy
tests of special relativity.

\bigskip
\centerline{\bf 2.1.  Building Lagrangians}
\medskip

We first construct all CPT-even Lorentz-violating
rotationally-invariant perturbations for a general renormalizable theory of
scalars, spinors and gauge mesons.  We also show that the matrix elements of
these interactions grow with energy more rapidly than those of the CPT-odd
Lorentz-violating interactions.

We begin by summarizing  some well-known properties of the Lorentz 
group~$SO(3,1)$\ 
\ref\rSW{{\it E.g.}, S. Weinberg, {\it The Quantum Theory of
Fields I}, \S 4.6 and \S 4.8, Cambridge U. Press, 1995.}.  
We assemble all the fields in the theory
into a big vector $\Phi$.
The action of
$\Lambda$, an element of $O(3,1)$, on these fields 
is effected by a unitary operator in
Hilbert space $U(\Lambda)$: 
\eqn\eIIi{U(\Lambda)^\dagger \Phi(x)U(\Lambda)=D(\Lambda)\Phi(\Lambda^{-1}x),}
where $D(\Lambda)$ is some finite-dimensional representation of the Lorentz
group.

The Lie algebra of $SO(3,1)$ may be written as the (complex) sum of two
commuting angular momenta, $\vec  J^{\,(+)}$ and $\vec J^{\,(-)}$.  An
irreducible finite-dimensional representation of the group 
may be labeled by two
half integers, $j_+$ and $j_-$, and is of dimension 
$(2j_++1)(2j_-+1)$.  We sometimes  write the fields forming the basis for
the irreducible representation as $\Phi_{m_+,m_-}$, where $m_\pm$ is the
eigenvalue of $J^{(\pm)}_z$.

The values of $(j_+,j_-)$ are $(0,0)$ for a scalar; $(0, 1/2)$ or $(1/2,
0)$ for a Weyl spinor, depending on its chirality; $(1/2, 1/2)$ for a
4-vector; $(1,1)$ for a traceless symmetric tensor; and the direct sum  of
$(1, 0)$ and $(0, 1)$ for an antisymmetric tensor. The complex conjugates
of a set of fields transforming  according to $(j_+,j_-)$ transform
according to $(j_-,j_+)$.

For $R(\vec e\, \theta)$, a rotation about an axis $\vec e$ by an
angle $\theta$, 
\eqn\eIIii{D\big(R(\vec e\, \theta)\big)=
\exp\big[-i(\vec  J^{(+)} + \vec  J^{(-)})\cdot
\vec e \,\theta\big].}
For $B(\vec e\, \phi)$, a Lorentz boost in a direction $\vec  e$ by
rapidity $\phi$, 
\eqn\eIIiii{D\big(B(\vec e\, \phi)\big)=
\exp \big[(\vec J^{(+)} - \vec  J^{(-)})\cdot \vec e \,\phi\big].}
Finally, for the anti-unitary CPT operator $\Omega$,
\eqn\eIIiv{\Omega^{-1}\Phi(x)\,\Omega = (-1)^{2j_+}\Phi^\dagger(-x).}

We are ready to begin our analysis.  From Eq.~\eIIii\ we note that
every rotationally invariant term ${\cal L}'$
in the Lagrangian  must lie in a
representation for which $j_+=j_- \equiv j$. Elementary
angular-momentum theory tells us that the 
term is: 
\eqn\eIIv{{\cal L}'\propto\sum_{m=-j}^j (-1)^m\Phi_{m,-m}.} 
If the state $|\psi\rangle$  is boosted 
in the $z$-direction by rapidity $\phi$, the
expectation value of ${\cal L}'$ is transformed according to:
\eqn\eIIvi{\langle\psi|U^\dagger\big(B(\vec e_z\phi)\big)\,{\cal
L}'(0)\,U\big(B(\vec e_z\phi)\big)|\psi \rangle \propto
e^{2j\phi}\langle\psi|\Phi_{j,-j}(0)|\psi\rangle + O(e^{(2j-2)\phi}).} 
That is,  $\langle {\cal L}'\rangle$ grows at large energy like
$E^{2j}$.  As we shall show shortly, the largest value of $j$ 
attainable
with operators of renormalizable type is $j=1$, 
a traceless symmetric
 tensor, which is CPT-even.  The only other rotationally-invariant
possibilities are $j=1/2$, a vector, which is CPT-odd, and $j=0$, a 
Lorentz invariant scalar.

Let us begin by considering only scalar fields.  With no loss of
generality we can consider these to be all real.  To attain $j=1$
we need at least two derivative operators, and for renormalizability we
can have no more than two (and no more than two scalar fields).
Thus the only possibility is
\eqn\eIIvii{\half\sum_{a,b}\partial_i\phi^a \epsilon_{ab}\partial^i\phi^b,}
where $\epsilon_{ab}$ is a real symmetric matrix and the sum runs over the
scalar fields. (Of course, we could just as well have said that the only
possibility is $\half\sum_{a,b}\partial_0\phi^a
\epsilon_{ab}\partial^0\phi^b$; the difference is a Lorentz invariant.)

Now let us consider spinors.  With no loss of generality we may take our
fundamental fields to be a set of $n$ $(1/2, 0)$ Weyl spinors, $u^a$; the
conjugate fields , $u^{a\dagger}$, are then $(0, 1/2)$ spinors. The only
Lorentz-invariant interactions either couple two spinors of the same type or
couple a spinor, a conjugate spinor, and a derivative. The most general free
Lagrangian is
\eqn\eIIviii{u^\dagger i(\partial_0-\vec\sigma\cdot\vec\partial)Zu +\half u^T
\sigma_y M u +\half u^\dagger \sigma_y M^\dagger u^*,}
where $Z$ is a Hermitian $n\times n$ matrix and $M$ is a (possibly complex)
symmetric matrix.  We can always make a linear transformation such that $Z$
is one and $M$ is real, positive, and diagonal.  We then have the theory of
$n$ free spin-1/2 particles, each of which is its own antiparticle.

The only way to construct a renormalizable interaction with $j=1$ is to
couple together a spinor, a conjugate spinor, and a derivative.   Thus we
obtain
\eqn\eIIix{\half\sum_{a,b}i\epsilon_{ab}u^{a\dagger}
\vec\sigma\cdot\vec\partial\,
u^b,}
where $\epsilon_{ab}$ is a Hermitian matrix.

A particularly simple case is that of two $(1/2, 0)$ Weyl spinors carrying
opposite charges under a $U(1)$ internal symmetry group.  Then one
Weyl spinor and the conjugate of the other can be joined to make a single
Dirac bispinor, $\psi$, the most general free Lagrangian is the standard
Dirac Lagrangian, and the most general $j=1$ interaction is
\eqn\eIIx{{\textstyle{1\over 4}}\bar\psi
i\vec\partial\cdot\vec\gamma[\epsilon_+(1+\gamma_5)+
\epsilon_-(1-\gamma_5)]\psi,}
where $\epsilon_\pm$ are real numbers.

We now turn to gauge mesons.  The couplings of the gauge mesons to the
scalars and spinors is completely determined by gauge invariance, so we need
only look at the gauge-field self-couplings.  Let us begin with a single
Abelian gauge field, and let us express it in terms of an electric and a
magnetic field, defined, as usual, by $E^i=F^{0i}$ and $B^i=\half
\epsilon^{ijk} F_{jk}$.  Out of these we can make three independent
rotational invariants of renormalizable 
type,\foot{The CPT-odd term $\vec A\cdot \vec B$, where $\vec A$ is the
vector potential, is discussed in the Appendix.}
which we may choose to be $\vec
E^2 -\vec B^2$, $ \vec E\cdot\vec B$, and $\vec B^2$.  The first two are
Lorentz invariants; we are left with $\vec B^2$, just as in the first example
of \S 1.

A general gauge group is locally the product of simple factors and Abelian
factors.  Gauge invariance forbids cross terms between gauge fields belonging
to different simple factors, so for each simple factor we have one
interaction, of the form $\sum_a \vec B^a\cdot \vec B^a$, where the sum runs
over the generators of the factor.   For the Abelian factors we can have
cross terms, but we can eliminate them by an orthogonal transformation on the
Abelian generators; thus we again obtain one $B^2$ term for each factor.

\bigskip

\centerline{\bf 2.2.  The Almost Lorentz-Invariant Standard Model}
\medskip

As an example, let us construct the $j=1$ interactions in the standard model
with three generations of quarks and leptons.  There are only a few terms in
the bosonic sector of the model: one $\vec B^2$ term for each factor
of the gauge group and one velocity-mixing term for the Higgs doublet, for a
total of four real parameters.

The number of parameters is much greater in the fermionic sector of the
model.
Each family of spinor fields transforms like the direct sum
of five inequivalent irreducible representations of the $SU(3)\otimes SU(2)
\otimes U(1)$ gauge group.  Gauge invariance excludes cross terms between
different irreducible representations, but not those between
families. Thus we have five $3\times 3$ Hermitian velocity-mixing
matrices of the form \eIIix, for 45 additional real parameters.

A small reduction in the number of parameters may be effected by field
redefinitions.   We can
rescale the space coordinates while leaving the time coordinate unchanged,
thereby 
eliminating one of the gauge-field $\vec B^2$ terms.  
Finally, we note that the minimal standard model conserves each lepton flavor.
Thus, we may multiply
the lepton fields by a phase factor depending
only on flavor so as to make two of the off-diagonal elements in one of the
lepton velocity-mixing matrices real while not affecting the
Lorentz-invariant Lagrangian. Thus
the standard model involves a total of 
$4+45-3=46$ independent Lorentz-violating (but CPT-conserving)
parameters. (We remark in passing that there are a comparable
number of 
parameters corresponding to $j=1/2$ CPT-violating departures from Lorentz
invariance.)

We used gauge invariance throughout this construction, but it has been
the gauge invariance of the classical Lagrangian.  Is this theory 
gauge-invariant as a quantum theory?  Phrased another way, we know that
anomalies cancel in the standard model, but does the cancellation persist
when we take account of the 46 noninvariant terms?  We demonstrate below that
it does.\foot{We were disappointed to discover this; we had hoped that the
condition of anomaly  cancellation would put some constraints on our 46 free
parameters.}

Our demonstration depends on the classic analysis of anomalies in
Lorentz-invariant non-Abelian gauge theories~\ref\rABZ{W. Bardeen,
Phys. Rev. 184 (1969) 1848; S. Adler and W. Bardeen, Phys. Rev. 182 (1969)
1517; W. Bardeen and B. Zumino, Nucl. Phys. B244 (1984) 421.}.
This begins by showing that the full theory is anomaly-free (to all
orders in perturbation theory) if the corresponding theory of massless
spinors coupled to external c-number gauge fields is anomaly-free.  If we
assemble all the spinor fields into a column vector $u$, the Lagrangian for
such a theory is:
\eqn\eIIxi{iu^\dagger (D_0-\vec\sigma\cdot\vec D)u,}
where the covariant derivative $D_\mu$ is defined by
$D_\mu=\partial_\mu+A_\mu$, with
$A_\mu$ a matrix-valued field composed of the gauge fields with
their associated coupling matrices.  If we define a gauge transformation
of the fields in the usual way,
\eqn\eIIxiii{\delta u=\delta\omega~u, \qquad \delta A_\mu=[\delta\omega,
A_\mu]-\partial_\mu\delta\omega\,,}
where $\delta\omega$ is an infinitesimal gauge transformation, it may
appear
that the theory defined by \eIIxi~ is gauge-invariant. However, this
is not necessarily so. A carefully regulated computation of $W(A)$,
the generating functional of connected Green's functions, yields
\eqn\eIIxiv{\delta W=-{1 \over {32\pi^2}}\,{\rm Tr} \int d^4x\,
\epsilon^{\mu\nu\lambda\sigma}\delta\omega
F_{\mu\nu}F_{\lambda\sigma},}
where $F_{\mu\nu}=\partial_\mu A_\nu-\partial_\nu A_\mu +[A_\mu, A_\nu]$.
Only if this expression (``the anomaly'') vanishes is the theory in fact
gauge-invariant.  Projection operators on irreducible representations of the
gauge group commute with $\delta\omega$ and $F_{\mu\nu}$; thus a convenient
way to evaluate the trace is to sum the contributions of the irreducible
multiplets.  For the standard model, this sum vanishes.

We wish to extend all this to a theory with a Lorentz-violating
interaction of the form ~\eIIix.  That is to say, \eIIxi~is replaced by
\eqn\eIIxv{iu^\dagger (D_0-(1-\half\epsilon)\,\vec\sigma\cdot\vec D)\,u,}
where $\epsilon$ is a Hermitian matrix acting only on the flavor indices.
This replacement alters the high-momentum behavior of the spinor propagator
and we must alter our regularization procedure accordingly.  If we use
massive regulator fields, the derivative terms in their Lagrangian must
be of the form \eIIxv, not \eIIxi. Likewise, if we regulate the measure in the
path integral, we must use the differential operator from \eIIxv, not 
from \eIIxi.

We must sum over irreducible multiplets, which may be chosen
to be eigenspaces of $\epsilon$. In each eigenspace, the Lagrangian is of the
form \eIIxv\ with $\epsilon$ a number.  We may introduce new variables
(denoted by primes) by:
\eqn\eIIxvi{\eqalign{ \vec x=(1-\half\epsilon)\,\vec{x}^{\,\prime}\,,
\qquad x_0=x'_0, \cr
\vec{A}^{\,\prime}=(1-\half\epsilon)\,\vec A,\qquad A_0'=A_0\,.\cr} }
In terms of these variables, the $\epsilon$ term disappears from
the Lagrangian (and from the regularization procedure).  The contribution of
each multiplet to the anomaly is the same as it would be if $\epsilon$
were zero, except that unprimed variables are replaced by primed
ones. This replacement has no effect on \eIIxiv, which is invariant under
general coordinate transformations.  Thus the contributions of the
irreducible multiplets are independent of $\epsilon$. If they cancel when
$\epsilon$ vanishes (as they do), they cancel for general $\epsilon$.

\bigskip
\centerline{\bf 2.3  From Lagrangians to Particles}
\medskip
Particle properties of free fields follow trivially from the
Lagrangian, but things are more difficult for interacting fields.
For simplicity we  study the energy-momentum relation  for  one
real scalar field of (renormalized) mass $m$. The generalization to
more complicated systems is straightforward. If the theory is
Lorentz-invariant, the inverse renormalized propagator has the form:
\eqn\eIIxvii{-iD^{-1}=(p^2-m^2)A(p^2),}
for some function $A$.  We normalize the field conventionally so that
$A(m^2)=1$, then add a Lorentz-violating interaction to the
theory with some small coefficient $\epsilon$, as in Eq.~\eIIvii.  We
begin in the linear approximation, retaining terms only 
first order in $\epsilon$. Later we investigate whether this  
approximation is
justified.  The addition to  $D^{-1}$ must
transform like the $00$ component  of a traceless symmetric tensor. The 
only possibility is a multiple of $4p^0p^0-g^{00}p^2=4\vec p^2+3p^2$.
The $p^2$ term can be absorbed in $A$, whence \eIIxvii~ becomes:
\eqn\eIIxviii{-iD^{-1}=(p^2-m^2)A(p^2) +\epsilon \vec{p}^{\,2} B(p^2),}
for some function $B$.  It is convenient to normalize the
Lorentz-violating interaction such that $B(m^2)=1$.

To
first order in $\epsilon$, the shift
in the zeroes of $D^{-1}$ is:  
\eqn\eIIxix{p^2=E^2-\vec{p}^{\,2}=m^2+\epsilon\vec{p}^{\,2}\,.}
The energy-momentum relation may be rewritten in the
seemingly conventional form
$E^2=\vec{p}^{\,2} c^2 + m^2c^4$, with $c$ the maximal attainable
velocity and $mc^2$ the rest energy of the particle. However, it must be
remembered that
$c^2$ has been changed by the factor to $1+\epsilon$ and $m^2$ has been
diminished by a factor of $(1+\epsilon)^2$.  Of course, the tiny mass shift
is of no experimental interest, but this is very much not the case for the
shift in $c^2$, as we saw in \S 1. Abrupt effects 
turn  on when the dimensionless parameter $\epsilon
\vec{p}^{\,2}/m^2$
is of order unity, while for gradual effects (like $\mu^+\rightarrow e^++
\gamma$) the energy at which 
the effect becomes significant can be many orders of magnitude smaller.

Even for gradual effects, $E$ is typically very large and we must ask
whether new Lorentz-invariant physics at high energies might affect our
predictions.  Eq.~\eIIxix\ shows that this does not happen.  Even though $E$ is
large, $p^2$ remains $O(m^2)$ and the possible new physics remains
irrelevant.

When can the linear approximation be trusted? It can certainly 
be trusted for
free field theory, where it is exact.  It can also be trusted for leptons and
electroweak gauge mesons.  For these particles, all couplings are weak, all
radiative corrections are small, and all propagators are well approximated by
those of free field theory.

Things are trickier for hadrons.  A detailed investigation of QCD
with Lorentz-noninvariant terms is beyond our ken, but we can make
an educated guess on the basis of a simple model.
Let's take QCD and rescale the  space coordinates but not the time
coordinate, as in Eq.~\eIIxvi.  In the new coordinates, we seem to have
a noninvariant theory, but it's just QCD wearing a false beard. All that
happens is that $p^2$ in Eq.~\eIIxvii\ is replaced by $p^2-\epsilon 
\vec p^{\,2}$.
Thus an expansion in powers of $\epsilon$ is in fact an
expansion in  powers of $\epsilon \vec{p}^{\,2}/\Lambda^2$, 
where $\Lambda$ is the
QCD mass scale.  Since typical hadron masses are $O(\Lambda)$, this implies
that for gradual effects the linear approximation is a very good one, while
for abrupt effects it is only a rough approximation.  Of course, a rough
approximation is not a useless one; it can give us a qualitative picture of
what  is going on, and even (with a modest amount of luck) yield
correct order-of-magnitude quantitative predictions.

\bigskip
\centerline{\bf 2.4 The Kinematics of Particle Decays}\medskip

In this section we analyze the decay of a particle into $n$ other
particles in our Lorentz non-invariant theories~\ref\rSLG{ Versions  of this
section appear in: \ 
S.L. Glashow, in Proc. TAUP-97 Nucl.Phys. B70 Proc. Suppl. (1999) 70; 
S. L. Glashow, in Proc. First Tropical Workshop, San Juan, Puerto Rico, ed.
Jos\'e F. Nieves, AIP Conf. Proc. 444 (1998) 119.}.
We make three
simplifying assumptions:  (1) that all  particles are spinless. The
extension to spin-${1\over 2}$  
particles is straightforward and has no effect on our
conclusions;  (2) that the  linear approximation is valid; and  (3) 
that the matrix elements of the invariance-violating
perturbation between particles with different masses are negligible.  Thus we
obtain a set of particles each  of which obeys an energy-momentum relation
of the form \eIIxix.  That is to say, the $a$th particle has, in addition to
its own mass, $m_a$, its own maximum attainable velocity (``its own velocity
of light'') $c_a$, and obeys the energy-momentum relation:
\eqn\eIIxx{E_a^2=\vec{p}_a^{\,2}c_a^2 + m_a^2 c_a^4.}
In what follows we use $a=0$ for the decaying particle and $a=1 \ldots n$
for the decay products.

A decay is kinematically permitted if we can arrange the decay products
such that their total momentum is $\vec p_0$  and
their total energy is
$E_0$.  Let $E_{\rm min}(\vec p_0)$ denote the  minimum total energy of the
decay products for given total momentum $\vec p_0$. The decay is  possible if
and only if:
\eqn\eIIxxi{E_0\geq E_{\rm min}(\vec p_0),}
because if
$E_{\rm min}<E_0$  we can obtain equality by adding
opposite transverse components to two of the decay momenta.

If we delete the transverse components of all decay momenta, we lower the
final-state energy without changing the total momentum.  
It follows that all momenta are collinear in
the configuration
of minimum total energy. We use this fact
to simplify our analysis and replace $\vec p_a$ by $p_a$, the longitudinal
(and only nonzero) component of $\vec p_a$.

$E_{\rm min}$ must be stationary  under variations of the decay momenta that
leave their sum unchanged. Introducing a Lagrange multiplier $u$, we must
make
\eqn\eIIxxii{\sum E_a -u\left(\sum p_a -p_0\right )}
stationary, where here and in the remainder of this section
the sum is over the decay
products. Differentiating with respect to  $p_a$, we find:
\eqn\eIIxxiii{u={dE_a\over dp_a} =v_a,}
where $v_a$ is the velocity of the $a$th particle.  (We've  used Hamilton's
equations at the last step.)  Thus all the decay  particles move with a
common velocity $u$. Furthermore. the relation:
\eqn\eIIxxiv{dE_{\rm min}=u\sum dp_a=udp_0\,,}
shows that $u=dE_{\rm min}/dp_0$.

We can now explore the limits of small and large $p_0$.
For $p_0=0$, the  minimum energy configuration is one in which all the decay
momenta vanish; the decay is  allowed if and only if:
\eqn\eIIxxv{m_0c_0^2\geq\sum m_ac_a^2\,.}
Of course, physics is certainly nearly Lorentz invariant, so that
the $c_a$'s can differ only very slightly. For all 
practical purposes, we can drop them from Eq.\ \eIIxxv.

For very  large $p_0$, $u$ must be ultrarelativistic and we can approximate
$E_a$ by $c_ap_a$.  The energy is minimized by giving all the momentum to
the particle with the smallest $c$. In this limit, the decay is allowed if:
\eqn\eIIxxvi{c_0 > \min_{a \neq 0} c_a\,,}
and  is forbidden if
$c_0< \min  c_a$.  (We must go
beyond the ultrarelativistic approximation to settle the question
if $c_0=\min c_a$.)

Eqs.~\eIIxxv~and~\eIIxxvi~are independent.  A decay is allowed or forbidden
in the high energy limit regardless of whether it  is allowed or forbidden 
at low energies.
It's interesting to ask what can happen at intermediate energies.  For
example, can there be alternating bands of allowed and forbidden energies?

We begin by answering this question 
when all the decay products have
nonzero masses; afterwards we'll extend our analysis to deal with massless
particles. 
Let's rewrite the condition for allowed decay,
$E_0 \ge E_{\rm min} $, as:
\eqn\eIIxxvii{m_0^2c_0^4 \geq E^2_{\rm min} (p_0) -c_0^2 p_0^2 \equiv Y(p_0).}
We shall prove  that every stationary point of $Y$ is a local maximum.
Differentiating \eIIxxvii\ yields:
\eqn\eIIxxviii{{{1\over 2}}{dY\over dp_0}=E_{\rm min}u-c_0^2p_0=
\sum(c_a^2-c_0^2)p_a,}
where we have used $E_a=c_a^2p_a/u$ at the last step.  
The derivative vanishes at
a stationary point,
so at least one term in the sum must be negative,
$c_0>\min_{a\ne 0} c_a$. Note that this condition coincides with
Eq.~\eIIxxvi, the condition that the process be allowed at high energy.

Differentiating once more, we find:
\eqn\eIIxxx{{1\over 2}{d^2Y\over dp_0^2}={du\over dp_0}\sum(c_a^2-c_0^2)
{dp_a\over du}={du\over dp_0} \sum (c_a^2-c_0^2){p_ac_a^2\over
u(c_a^2-u^2)}.}
At a stationary point, we can use the vanishing of \eIIxxviii~to write this
as:
\eqn\eIIxxxi{\eqalign{{1\over 2}{d^2Y\over dp_0^2}=&{du\over dp_0}\sum
(c_a^2-c_0^2) {p_a\over u}\left [{{c_a^2}\over {c_a^2-u^2}}-{c_0^2\over
c_0^2-u^2}\right ]\cr =&-u{du\over dp_0}\sum {p_a(c_a^2-c_0^2)^2\over
(c_a^2-u^2)(c_0^2-u^2)}\,.\cr}}
As $p_0$ increases monotonically from zero
to infinity,
$u$ increases monotonically from
zero to $\min c_a$.   Thus $du/dp_0$ is positive, as 
are the factors $(c_a^2-u^2)$ and
$(c_0^2-u^2)$, these last by ~\eIIxxvi.  This completes the  proof.

Because every stationary  point of $Y$ is a local maximum, $Y$ can have at most
one stationary point.  If $Y$ has no stationary points, it is a monotone
function of $p_0$ and can cross $m_0^2c_0^4$ at most once.  This leads to
four possibilities:  (1) The decay is allowed at all energies.\ \   (2) The
decay is
forbidden at all energies.  (3) The decay is allowed for all energies below a
certain energy and forbidden for greater energies.  (4) 
The decay is forbidden for
all energies below a certain energy and allowed for all greater energies. If
$Y$ has one stationary point,\foot{The necessary and sufficient conditions 
for $Y$ to have an extremum 
are Eq.\ \eIIxxvi\  and \ $\sum(c_a^2-c_0^2)\,m_a >0\,$ (which
ensures that $Y(p_0)$ increases near $p_0=0$).}
 it may cross $m_0^2c_0^4$ twice, the first time
from below and the second time from above. This adds one more possibility: (5)
The decay is forbidden for a certain band of energy but allowed for all
energies above or below this band.
There are no other possibilities.

Our arguments break down if there are massless particles among the decay
products.
(For example, we can no longer write energy and momentum as functions  of
velocity.)  Nevertheless, we now show that our conclusions remain valid.

If there is more than one massless particle among the decay products, we can
lower $E_{\rm min}$ by 
giving all the momentum carried by the massless particles
to the one with the smallest value of $c$.  Thus no 
generality  is lost 
by restricting ourselves to the case in which there is only one
massless particle, which we label by $a=1$.

Consider a configuration with $p_1$ fixed and the remaining
momentum, $p_0-p_1$, distributed among the massive particles so as to
minimize their total energy.   The total energy of this configuration is:
\eqn\eIIxxxii{E=c_1 p_1+ E'_{\rm min}(p_0-p_1)\,,}
where $E'_{\rm min}$ is the minimum energy computed for the massive particles
only. Thus,
\eqn\eIIxxxiii{dE/dp_1 = c_1-u'(p_0-p_1),}
where $u'$, the common velocity  of the massive particles, is a monotone
increasing function of their total momentum.  There are two cases:
(1) $u'(p_0)< c_1$, so that $E$ is a monotone increasing
function of $p_1$ and $E_{\rm min}=E'_{\rm min}(p_0)$.  The massless
particle carries no momentum so that the analysis is the same as
in the massive case. \  (2) $u'(p_0)>c_1$, which is 
possible only if:
\eqn\eIIxxxiv{c_1 < \min_{a \neq 1} c_a\,,}
for otherwise $u'$ cannot reach $c_1$.  In this case,
increasing $p_1$ lowers $E$ until $u'$ reaches $c_1$, and
\eqn\eIIxxxv{E_{\rm min}=c_1(p_0-p_0')+E_{\rm min}'(p_0'),}
where $p_0'$ is defined by $u'(p_0')=c_1$.  This analysis proceeds as in
the massive case up to and including Eqs.~\eIIxxviii~and~\eIIxxvi, but the
computation of $d^2Y/dp_0^2$ is different.  Because only $p_1$
depends on $p_0$,
\eqn\eIIxxxvi{{1\over 2}{d^2Y\over dp_0^2}=(c_1^2-c_0^2){dp_1\over dp_0}
=(c_1^2-c_0^2)\,c_1 <0,}
by Eqs.~\eIIxxvi~and~\eIIxxxiv.  Just as in the case of only massive
decay products, every stationary  point of $Y$ is a local maximum.
Thus we
reach the same conclusions here as we did there.\foot{The borderline
situation, where 
the stationary  point of $Y$  occurs at $p_0=p_0'$,
must be treated separately.
In this case we must use the analysis for $p_0<p_0'$ to compute $d^2Y/dp_0^2$
from below and that for $p_0>p_0'$ to compute it from  above. The two answers
don't agree because the second derivative of $Y$ is not continuous at the
stationary point, but it doesn't matter because both are negative.}

\newsec{Applications}
Here we examine effects of CPT-conserving departures from Lorentz
invariance on the behavior of ultra-relativistic particles. 
Our treatment of leptonic phenomena (in \S 3.1 and \S3.2) is not subject
to the qualifications discussed in \S 1.4. However, the
hadronic phenomena (in \S 3.3) often involve hadron 
energies so large that $E
\sqrt{\delta c^2}$ is comparable to   the QCD mass scale. Thus, our
treatment of the $c_a$ as energy-independent parameters is not  always
justified.  This error  affects the values of the Lorentz-violating parameters
needed to produce novel phenomena, but not the phenomena themselves.

\bigskip
\centerline{\bf 3.1. Phenomena Involving Charged Leptons}\medskip

{\it Tests of Lorentz Invariance from Photon Stability: }
Suppose that the velocity of
light exceeds the maximal attainable velocity (MAV) of electrons, {\it i.e.,}
$c_\gamma>c_e$, where for the
moment we ignore
possible flavor and helicity dependences. It follows~\rCG~that photons of
sufficient energy are unstable. In particular, the decay: 
\eqn\egammadecay{\gamma\rightarrow e^-+e^+}
becomes kinematically permitted when  the photon
energy $E$ exceeds $2m_e/\sqrt{\delta_{\gamma e}}$, where $\delta_{\gamma
e}\equiv c_\gamma^2-c_e^2$.  
The decay rate of photons well above this threshold is $\Gamma_{ee}\simeq
{1\over 2} \alpha \delta_{\gamma e} E$. The fact that
primary cosmic-ray photons with energies up to 20~TeV have been
detected lets us set the limit $c_\gamma-c_e < 10^{-15}$.
Note that $\vert \delta_{\gamma e}\vert $ is not so well constrained:
the absence of vacuum \v Cerenkov radiation by electrons
with energies up to 500~GeV
sets the  limit: $c_e-c_\gamma < 5\times 10^{-13}$. In a similar manner,
much weaker constraints may be placed on the 
MAVs of the heavier charged leptons.

More stringent tests of Lorentz invariance might be imagined to
result from the stability of photons under decay into two neutrinos, 
for
which the threshold energy depends on
tiny neutrino masses. This mode is forbidden in
the standard model, which attributes neither masses nor  magnetic moments
to neutrinos, but there is a considerable body of empirical evidence for
neutrino oscillations, and hence for neutrino masses. 
The non-standard physics responsible for
neutrino masses could generate  neutrino magnetic moments
via loop diagrams, thereby enabling
the decays $\gamma\rightarrow \nu+\nu'$ and $\bar \nu +\bar \nu'$ 
(with $\nu$ and $\nu'$ necessarily
distinct chiral neutrinos).

The decay rate of a photon with energy
$E$ well above threshold into neutrinos is:  \
 $\Gamma_{\nu\nu'}\sim \lambda^2 \mu_B^2
\,\delta_{\gamma\nu}^2 E^3$, where
the flavor-changing ``magnetic moment'' of the neutrino
is $\lambda \mu_B$ with $\mu_B\equiv
e/2m_e$ (the Bohr magneton). We 
assume $c_\nu=c_{\nu'}$ and put $\delta_{\gamma\nu}= c_\gamma^2-c_\nu^2$. 
Because the magnetic interaction is
dimension-five (rather than dimension-four, like the electric),
$\Gamma_{\nu\nu'}$
is {\it quadratic\/} in 
$\delta_{\gamma\nu}$ rather than linear.\foot{Our estimates of $\Gamma_{ee}$
and $\Gamma_{\nu\nu'}$ are lowest order in the 
appropriate Lorentz-invariant operators,
with photons satisfying  the Lorentz-violating dispersion relation
$E^2-{\vec p}^{\, 2}= \delta E^2\equiv$``$m_\gamma^2$''. 
The decay rates are $\Gamma_{ee}\big\vert_{\rm
rest}\sim \alpha
m_\gamma$ and $\Gamma_{\nu\nu'}\big\vert_{\rm rest}
\sim (\lambda\mu_B)^2 m_\gamma^3$ in the ``photon
rest frame.'' 
 These expressions yield our results
 when boosted
to the lab frame.} It follows that the range of an energetic photon is
cosmological even if the neutrino magnetic moment is (implausibly)
set equal to its experimental upper limit~\ref\ederb{A.I. Derbin {\it et
al.,}  Phys. At. Nucl. 57 (1994) 222.}: 
$\lambda=2\times10^{-10}$. Consequently, no strong bound on
$c_\gamma- c_\nu$ can be deduced from observations of energetic
cosmic-ray photons.
\medskip

{\it Radiative Muon Decay? }  The decay mode mode $\mu\rightarrow e
+\gamma$  is often searched for  but never found. This is not surprising:
it is forbidden in the minimal standard model. Although 
induced by radiative corrections  in models with neutrino masses, the
expected branching ratio is far too small to be detected.  However, and as
we noted elsewhere~\ref\rmudecay{S. Coleman and S.L. Glashow, HUTP-98/A076
}~\ref\rsuz{ S. Coleman and S.L. Glashow,  HUTP-98/A080, to appear in Proc.
Neutrino '98, Takayama, Japan.}, Lorentz-violating  perturbations  that are
not flavor diagonal can lift the accidental symmetry  ordinarily preventing
radiative decay. 

Recall that the velocity eigenstates of high-energy leptons do not in
general coincide with their  mass eigenstates at low energy.  
In the following analysis, we ignore possible mixings of electrons and
muons with tau leptons. The relevant portion of the Lagrangian
takes the following form in the preferred frame:
\eqn\elag{ 
\pmatrix{\bar\mu&\bar e\cr}\,\vec{\gamma}\cdot
(\vec{p}-e\vec{A})\,
\left\{{\cal C}_R \,{\textstyle{1\over2}}(1+\gamma_5) + 
{\cal C}_L\, {\textstyle{1\over2}}(1-\gamma_5)\right\}\,
\pmatrix{\mu\cr e\cr}\,,}
where $\mu$ and $e$ denote fields corresponding to mass eigenstates.
The matrices ${\cal C}_{L,R}$ 
(which would be unit matrices were Lorentz symmetry unbroken)
are:
\eqn\ematrix{{\cal C}_a\equiv {\textstyle{1\over2}}\,
\pmatrix{2\bar{c}_a+\delta c_a
\cos{2\theta}_a&\delta c_a\sin{2\theta_a}\cr
\delta c_a\sin{2\theta_a}&2\bar{c}_a-\delta c_a\cos{2\theta_a}\cr}\;,  
\quad a=L,\,R\,.} 
The mixing angle $\theta_{L}$ 
determines the velocity eigenstates  of left-handed
leptons (or right-handed antileptons) whose MAVs
are $\bar{c}_L\pm {1\over2}\delta c_L$.
Similarly, $\theta_R$ determines the
velocity eigenstates of right-handed leptons (or left-handed antileptons),
whose MAVs differ by $\delta c_R$. 
All four maximal velocities are known to be very close to $c_\gamma$, the
velocity of light in vacuum.  

Electroweak gauge invariance implies that the matrix ${\cal C}_L$
appears in the kinetic-energy of neutrinos as well as charged leptons.
In \S 3.2 
we show how the parameters $\bar{c}_L$, $\delta c_L$ and $\theta_L$  
 may be constrained by experiments or observations involving neutrinos. 

For the analysis to follow, it is convenient to define the small parameters:
\eqn\eparam{\epsilon_a^2\equiv \vert \delta c_a\,\sin{2\theta_a}\vert^2\,,
\qquad a=L,\,R\,.}
The flavor-changing terms in Eq.~\elag, namely
\eqn\eterm{{\textstyle{1\over2}}\,(\epsilon_R+\epsilon_R)\, \bar\mu\,
\vec\gamma\cdot(\vec p-e\vec A)\,e +
{\textstyle{1\over2}}\,(\epsilon_R-\epsilon_R)\, \bar\mu\,
\vec\gamma\cdot(\vec p-e\vec A)\,\gamma_5\,e +\rm h.c.,}
induce the decay process $\mu\rightarrow e+\gamma$. 
We shall see that its 
helicity-dependent partial decay rate is $\sim\!\alpha\epsilon_a^2
M\gamma^3$, where $M$ is the muon mass and $\gamma$ is its Lorentz factor
in the preferred frame. This rate
increases with the cube of the muon energy rather
than falling with $1/E$.  Thus
flavor-changing Lorentz-violations, if they are present, will cause
$\mu\rightarrow e+\gamma$ to become the
dominant decay mode of muons at sufficiently high energies.

The rate of radiative muon decay is controlled 
by the muon energy and the magnitudes of
the Lorentz-violating parameters. Otherwise, it is a first-order
electromagnetic effect, not a weak decay. Its branching ratio is $B\sim 
\alpha\epsilon_a^2 \gamma^4 M\tau_0$, where $\tau_0$ is the lifetime of a
muon at rest.  Departures from Lorentz invariance also modify the  
rates of allowed processes, such as $\pi\rightarrow \mu +\nu$ and
$\mu\rightarrow e+\nu+\bar\nu$,  but in these cases  the conventional decay
rates $\Gamma_0/\gamma$  and their Lorentz-violating  corrections (of the
form  $\epsilon^2 \gamma^3 \Gamma_0$) are  both intrinsically weak. They do
not involve the enormous  enhancement factor $\alpha M\tau_0\simeq
2.6\times 10^{15}$ that appears in the  branching ratio for radiative muon
decay. That's why the most sensitive tests of Lorentz invariance in this
context are obtained from the study of muons, and in particular, from the
search for a lifetime anomaly of muons at ultra-high energies.

The interaction \eterm, treated in lowest order perturbation theory, yields
the rate for radiative muon decay.
A straightforward but tedious computation
(generously carried out for us by Mark Wise) 
 yields the following result for its branching ratio
{\it when the muon is  at rest:}
\eqn\erest{B= {\alpha M \tau_0\over 4}
\big(\epsilon_L^2+\epsilon_R^2\big)\simeq 6.4\times10^{14}
\,\big(\epsilon_L^2+\epsilon_R^2\big)\,. }
The current experimental limit~\ref\rmu{R.D. Bolton, Phys. Rev  D38 (1988)
2077.},
$B<4.9\times10^{-11}$,  yields an
upper limit on the relevant Lorentz-violating parameter:
\eqn\edar{\epsilon_L^2+\epsilon_R^2< 
8\times 10^{-26}\qquad\rm from\ muon\ decay\ at\ rest.}

The branching ratio for Lorentz-violating radiative muon decay is a rapidly
increasing function of the energy. 
Direct searches for it do 
not seem 
feasible. Nonetheless one might detect the onset of this mode through
its effect on the muon lifetime, which for
 ultrarelativistic 
 left-handed $\mu^-$ (or right-handed $\mu^+$) is:
\eqn\erel{\tau_L(\gamma)= {\gamma\tau_0\over 1+ b_L\gamma^4}\,,\quad{\rm
where}\quad b_L\equiv {\alpha M \tau_0\over 30}\,(68\,\epsilon_R^2
+\epsilon_L^2)\,,}
and for 
ultrarelativistic right-handed $\mu^-$ (or left-handed $\mu^+$) is:
\eqn\erell{\tau_R(\gamma)= {\gamma\tau_0\over 1+ b_R\gamma^4}\,,\quad{\rm
where}\quad b_R\equiv {\alpha M \tau_0\over 30}\,(68\,\epsilon_L^2
+\epsilon_R^2)\,.}
(These calculations were also carried out for us by Mark Wise.)
At sufficiently high energy, the lifetime of muons with either helicity
decreases with $\gamma^{-3}$
rather than increasing with $\gamma$.

The CERN $g-2$ experiment, aside from measuring the muon's anomalous magnetic
moment, offers a precise test of the energy dependence of
its lifetime. At $\gamma=29.3$ (corresponding to the `magic energy' at
which the experiment was performed) the results
confirm the expected time-dilation of thed muon lifetime to
an accuracy of one part in a thousand~\ref\g{{\it E.g.,}
F. Combley, F. J. M.  Farley and
E. Picasso, Phys. Rep.  68 (1981) 93.}. Because the muons in the ring favor
neither helicity when time-averaged,
we obtain the limit $b_L+b_R<2.7\times 10^{-9}$, or:
\eqn\egmt{\epsilon_L^2+\epsilon_R^2< 5\times 10^{-25}\qquad{\rm from\ muon\ \;}
g-2\,.}
which is inferior to that obtained from the direct
search, Eq.\ \edar, but not by much!  (The 
agreement between
theoretical and experimental values of $g-2$ provides 
an independent but much weaker  test of Lorentz invariance.) 

\medskip
{\it  The Muon Collider---A Threat Averted! }
Lorentz-violating effects consistent with the above constraints  could
interfere with the operation of
a future muon collider. In \rmudecay\ and
\rsuz, we took 
as a necessary
condition for its proper functioning, that the muon decay rate at the design
energy  of the collider
must not exceed twice its expected value. This criterion, assuming
unpolarized beams, translates to  ${1\over 2}(b_L+b_R)\gamma^4<1$. Thus a
1~TeV collider requires  $b_L+b_R< 2\times 10^{-16}$ (or
$\epsilon_L^2+\epsilon_R^2 <3\times 10^{-32}$).
The constraints discussed above are much weaker than this.
However, we suggested that
constraints sufficient to protect the muon collider
might be obtained through studies of cosmic rays. 

The
highest energy cosmic-ray muons arise from forward decays of secondary
pions. These are mostly longitudinally polarised $\mu^-_R$ and $\mu^+_L$,
so that searches for a lifetime anomaly  of these muons  can
set a bound on $b_R$, which is  
proportional to the linear
combination $68\,\epsilon_L^2+\epsilon_R^2$.  Although $\epsilon_L$ 
can be severely constrained by  observations of neutrino
oscillations, it's  possible that
$\epsilon_R\gg \epsilon_L$.  Absent any {\it a
priori\/} knowledge of the ratio $\epsilon_L/\epsilon_R$, a bound on $b_R$
yields a 69-fold weaker bound on $b_R+b_L$

In particular,
we suggested~\rmudecay\  how a comparison of underground
muon fluxes with different flight times but the same slant depth 
in rock might
yield the required constraint. Following our suggestion,
V.S. Narasimham and M.R.
Krishnaswamy~\ref\rks{As reported to us by B.V. Sreekantan, private
communication.} have analyzed data taken at the Kolar Gold fields. Their
preliminary result is $b_R<8\times 10^{-18}$.   
An even stronger constraint
has been obtained by R. Cowsik and B.V. Sreekantan~\ref\rcs{R. Cowsik and
B.V. Sreekantan, {\tt hep-ph/9811241}
(to appear in Physics Letters B), and private communication.} from
their considerations
of data concerning horizontal air showers. They obtain the remarkably
strong  bound
$b_R<10^{-25}$. From Eq.~\erell, this bound corresponds to:
\eqn\aira{68\,\epsilon_L^2+\epsilon_R^2<10^{-39}\qquad \rm from\ 
air\ showers\,.}
The results to either of these
recent cosmic-ray analyses are sufficient to safeguard
the muon collider from a weakening of the relativistic dilation of the muon
lifetime.
 Indeed, taken by
themselves, they provide sensitive tests of special relativity. For
example, the air shower result (with $\epsilon_L=\epsilon_R$
and maximal velocity mixing of muons and electrons) 
yields the following  constraint 
on the difference  of maximal velocities of the two velocity eigenstates:
\eqn\eairb{\big\vert c'-c\big\vert_{e\mu} <4\times
10^{-21}\,.}
We shall show in \S 3.2 how neutrino physics can do even
better.
\vfill\eject

\centerline{\bf 3.2 Phenomena Involving Neutrinos}\medskip

Although {\it differences\/}  among neutrino velocities are severely
constrained directly (by studies of neutrino oscillations, discussed
below)  and indirectly (from constraints involving muons or electrons
discussed in \S 3.1 via $SU(2)$ gauge invariance),  there are only weak
constraints on neutrino maximal velocities as such. The current limit, $
|c_\nu-c_\gamma| < 10^{-8}$,  results from the detection of neutrinos
from supernova~1987a~\ref\rsto{L. Stodolsky, Phys. Lett.  B201 (1988)
353.}. A stronger limit, $|c_\nu-c_\gamma| < 2\times 10^{-16}$, may be
set if neutrinos from gamma-ray bursts at cosmological distances
could be detected~\ref\jb{E. Waxman and J. Bahcall, Phys. Rev. Lett. 
78  (1997) 2292.}, but this result pales in comparison to 
other tests of Lorentz invariance that have been (or could be) set.
\medskip

{\it Lorentz-Violating Neutrino Oscillations? }
We showed in \rCG\ and \rsuz\ how CPT-conserving
Lorentz violations lead to neutrino
oscillations even if neutrinos are massless.\foot{Similar effects can
result from violations of the equivalence principle rather than special 
relativity~\ref\rvep{M. Gasperini, Phys. Rev. D38 (1988) 2635\semi A. Halprin
and C.N. Leung, Phys. Rev. Lett. 67 (1991) 1833.}.} However,
observable neutrino oscillations  may result from a combination of effects
involving neutrino masses and Lorentz
violation.

Neutrinos with modest energy (even solar neutrinos) 
are nevertheless ultra-relativistic particles because neutrino masses are
known to be small.  For this reason, searches for neutrino oscillations can
provide  exquisitely sensitive tests of Lorentz invariance.  We assume
there exist three chiral neutrinos with Majorana masses  given by the
complex symmetric matrix $m$ in a flavor-diagonal basis. Conventional
neutrino oscillations are described in terms of the Hermitian squared-mass
matrix $m^2=mm^\dagger$. In particular, observable oscillation effects
depend on two differences of squared masses and four parameters akin to the
Kobayashi-Maskawa angles and phase in the quark sector.

With Lorentz symmetry violated, the  description of neutrino oscillations
becomes much more complicated.
Although we usually neglect CPT-violating
interactions, in this case it is hardly any work to include them.  Thus, in
addition to our usual velocity-mixing term, parameterized by c, a $3\times
3$ Hermitian matrix of maximum attainable velocities, we allow
the most general CPT-odd symmetry violating interaction, $u^\dagger b u$,
where b is also a $3\times 3$ Hermitian matrix.  
The energies of ultra-relativistic neutrinos with definite momentum $p$ are
the eigenvalues of the matrix:  
\eqn\enudisp{cp + m^2/2p + b\,.}
Neutrino energy eigenstates in the limit of high energy are the
eigenvectors of 
the matrix $c$, just as at low energy they are the eigenvectors
of the matrix $m^2$.

To avoid undue complexity, 
we limit ourselves to a discussion of two-flavor neutrino
oscillations. Imagine neutrinos to be produced with
a definite momentum and flavor ( $\nu_\ell$,
where $\ell= e$ or $ \mu$ ) and detected after
travelling a distance $R$ through empty space. Their oscillations satisfy a
seemingly conventional formula:
\eqn\enuosc{ P(\nu_\ell\rightarrow \nu_\ell)= 1- \sin^2{2\Theta }
\sin^2{\left\{\Delta R/4\right\}}\,,}
where the mixing angle $\Theta$ and phase factor $\Delta$ 
appearing in Eq.~\enuosc\ (and expressed in the flavor basis) are given
implicitly in terms of eight convention-independent parameters: 
\eqn\emess{\eqalign{ 
&\Delta \sin{2\Theta} = \delta m^2 \sin{2\theta_m}/E +
2\delta b\,e^{i\eta}\sin{2\theta_b} + 
2 \delta c\,e^{i\eta^\prime}E\sin{2\theta_c} \,,\cr
&\Delta \cos{2\Theta} = \delta m^2 \cos{2\theta_m}/E +
2 \delta b\,\cos{2\theta_b} + 2 \delta c\,E\cos{2\theta_c}
\,.\cr} }
The parameters characterizing the oscillations
are:  
three mixing angles, $\theta_m$, $\theta_b$ and $\theta_c$,
two complex phases, $\eta$ and $\eta'$,
and the differences between the eigenvalues of the
matrices $m^2$, $b$, and $c$, denoted respectively by $\delta m^2$, 
$\delta b$ and $\delta c$. (Note that $\delta c$ here is the
same as $\delta c_L$ in \S 3.1.) To 
illustrate the possibilities inherent
in Eqs.~\enuosc\ and \emess, we mention a few
special cases of Lorentz-violating two-flavor neutrino oscillations:
\eqna\exam
$$\eqalignno{
P(\nu_\ell \rightarrow \nu_\ell) \, & =1-
{\sin^2{\left\{(\delta m^2R/4E)\sqrt{1+(E/E_0)^2}\right\}} \over
1+ (E/E_0)^4} & \exam a\cr
& =1- {\sin^2{\left\{(\delta v\,RE/2)\sqrt{1+(E_0/E)^2}\right\}} \over
1+ (E_0/E)^4} & \exam b\cr
& = 1-  \sin^2{2\theta}\sin^2{\left\{R(\delta m^2/4E 
+\delta b/2 + \delta c E/2)
\right\}} & \exam c\cr}$$
where $E_0\equiv \delta m^2/(2\delta c)$.  Eq. \exam{a}\ corresponds to
$\theta_m=\pi/4$ with 
 $\delta b=\sin{2\theta_c}=0$. It yields maximal mass oscillations for
 $E\ll E_0$, but essentially none for $E\gg E_0$.
Eq. \exam{b}\ corresponds to a converse case with
$\theta_c=\pi/4$ and
 $\delta b=\sin{2\theta_m}=0$: maximal velocity oscillations 
at high energy, but none at
low energy.

To obtain Eq. \exam{c}, we set all three mixing angles equal and put
$\eta=\eta'=0$. In this case, the energy-momentum eigenstates are
independent of energy.
This example encompasses all  three scenarios
discussed by Foot, Leung and Yasuda~\ref\rfoot{R. Foot, C.N. Leung, and O.
Yasuda, {\tt hep-ph/9809458}.}~for atmospheric neutrino oscillations --- 
each of which, they say, is consistent with
the 
atmospheric neutrino data reported by Super-Kamiokande~\ref\superk{T. Kajita
(for the SuperKamiokande Collaboration), to appear in 
Proc. Neutrino '98, Takayama, Japan.}.

Conventional neutrino oscillations depend only on $R/E$, the ratio of the 
flight-length of the neutrino to its energy. If Lorentz symmetry is
violated, the dependence on these parameters is more complicated.
Nonetheless, neutrino experiments performed at a variety of energies can
severely constrain the Lorentz-violating parameters. Let's give a simple
example related to accelerator searches for $\nu_\mu$--$\nu_e$
oscillations. The strongest current limit on $\delta m^2$ (with
$\sin^2{2\theta_m}\sim 1$), $\delta m^2<0.09$~eV$^2$, follows  from a
relatively low energy experiment~\ref\rang{C. Angelini {\it et al.,} Phys.
Lett. B179 (1986) 307.}. Higher energy neutrino experiments,  such as
\ref\bruck{E.B. Brucker {\it et al.,} Phys. Rev. D34 (1986) 2183.},  offer
less stringent constraints on $\delta m^2$ but are better suited to search
for Lorentz-violating velocity oscillations. 
From that experiment,
and assuming $\sin^2{\theta_m}\sim 1$,
we find for the difference of the maximal velocities
of the two velocity eigenstates: 
\eqn\enuvel{\big\vert c' - c\big\vert_{\nu_e\nu_\mu} < 6\times 10^{-22}\,.}

Finally, we note that stringent  constraints on the CPT-violating
parameters that affect neutrino oscillations have been obtained from
altogether different laboratory experiments.
According to one of its collaborators~\ref\hunt{L.H. Hunter,
private communication.},
the spectroscopic test of Lorentz invariance
described in \rberg\ constrains the 
parameter $b_3$ (as defined by Colladay and
Kostaleck\'y~\rkost). They obtain $\vert b_3\vert < 7\times 10^{-19}$~eV for
electrons and  $\vert b_3\vert < 1.2\times 10^{-21}$~eV for nucleons. The
former result, expressed in our model and in the preferred frame, corresponds
loosely\foot{Here we neglect possible CPT violations involving right-handed
electrons.}
to the constraint $\vert b_{ee}\vert < 3\times 10^{-16}$~eV.  
This result suggests that CPT-violating effects are too small to 
affect neutrino oscillations, except when the source distance far exceeds
the diameter of the Earth (as in the case of solar 
or extra-solar neutrinos).

\bigskip

\centerline{\bf 3.3 Phenomena Involving Hadrons} \medskip

To each particle species we assign a maximal attainable velocity $c_a$.
That is, we assume that a dispersion relation of the form $E^2=c_a^2 p^2
+m_a^2c_a^4$ describes a particle of type $a$  moving freely in the
preferred frame. Many  Lorentz-violating (but CPT-conserving) phenomena may
be described in terms of the purely  kinematic effects of these parameters.
For simplicity, we ignore the helicity dependence of the
MAVs, although it could easily be taken into account. 
Flavor-changing effects are not relevant to the phenomena discussed in this
section and are likewise ignored. \medskip

{\it The Neutral Kaon System: } 
Lorentz-violating effects can  be abrupt or gradual. We
gave examples of each in \S 1---the  sudden onset of vacuum \v Cerenkov
radiation by energetic protons, and an
energy-dependent modulation of the behavior of neutral kaons. Although our
primary focus here is on abrupt hadronic effects, it is illustrative
to examine the latter phenomenon in more detail. 
We consider the special case in which the velocity and mass eigenstates of
neutral kaons coincide, and in which
the MAVs of $K_L$ and $K_S$ are not the same:
$ c_{K_L}-c_{K_S} \ne 0$.  
This leads to an energy dependence of their apparent  mass
difference, as determined by observations of   time-dependent interference
phenomena:   
\eqn\ekaons{\Delta M = \Delta M\big\vert_{\rm rest} +
M\beta\gamma^2(c_{K_L}-c_{K_S})\,,}
where $\gamma$ and $\beta$ are the Lorentz factor and velocity of the
decaying kaons.
A careful analysis of the experimental data carried out by Hambye, Mann and
Sarkar~\ref\rsark{T. Hambye, R.B. Mann and U. Sarkar, Phys. Lett.
B421 (1998) 105.}~yields the bound:\foot{Other tests of CPT and Lorentz
invariance using neutral mesons are discussed in \rsark\ and elsewhere
\ref\rkaons{V. A.
Kosteleck\'y, Phys. Rev. Lett. 80 (1998) 1818 and Indiana University
preprint IUHET 396 {\tt hep-ph/9810352}.}.}
\eqn\ekaons{\vert
c_{K_L}-c_{K_S}\vert < 3\times 10^{-21}\,.}

\medskip

{\it Stable Neutral Pions? }Let's  turn to abrupt phenomena. A simple example
involves neutral pions and photons. Suppose that $c_\gamma- c_{\pi^0} > 0$.
The process $\pi^0\rightarrow 2\gamma$ (the dominant decay mode of neutral
pions) becomes kinematically forbidden for pions with energies exceeding 
$E=m_\pi\big/\sqrt{c^2_\gamma- c^2_\pi}$. Conversely, photons with energies
significantly above $E$ will decay 
rapidly according to the scheme: $\gamma\rightarrow \gamma
+ \pi^0$. This example is not entirely academic. Suppose, for example:
\eqn\establepi{c_\gamma- c_{\pi^0} =c_\gamma-c_e =
 c_\gamma- c_{\pi^0}=10^{-22}\,.}
 For this
case, all modes of $\pi^0$ decay are kinematically forbidden at pion
energies exceeding $10^{19}$~eV. Thus ultra-high-energy (UHE)
primary cosmic rays may include
neutral pions (if they are stabilized by tiny departures from Lorentz
invariance), but not photons 
(which would be destabilized by the same mechanism).
\medskip

{\it Stable Neutrons? }
Ordinarily, neutron decay ($n\rightarrow p+ e^-+\bar \nu$) is
allowed but proton beta decay ($p\rightarrow e^+ + n+ \nu$) is kinematically
forbidden.  
As we have seen in \S~2.4,
departures from Lorentz invariance can affect the
kinematics of decay processes. They even can invert 
this pattern! To see how this can come about, we examine the case
$c_p=c_e=c_\nu < c_n$. Conventional relativistic kinematics
may be used in this example (with
$c_p$ as ``the speed of light''), provided that the
neutron is assigned an effective mass $m_{\rm eff}$  given by:
\eqn\emeff{ m^2_{\rm eff} \equiv m_n^2-(c_p^2-c_n^2)\,\vec p^{\,2}\,,}
where $\vec p$ is its momentum in the preferred frame. 
Neutron beta-decay is allowed if and only if
$m_{\rm eff}> m_p+m_e$. Expressed in terms of
the neutron energy $E$ in the preferred frame, this condition
becomes:
\eqn\enbeta{ E<E_1=\sqrt{{m_n^2 -(m_p+m_e)^2\over c_p^2-c_n^2}}
\simeq 
2.7\times 10^{19}\; \left[{10^{-24}\over c_p-c_n}\right]^{1/2}\; \rm eV.}
With our choice of Lorentz-violating parameters,
{\it neutrons with energies exceeding $E_1$ are 
stable particles that can
be present among 
UHE cosmic rays.}

In a similar manner, we can deduce the necessary and sufficient
condition for proton beta decay to be kinematically
permitted. It is:
\eqn\epbeta{ E>E_2\simeq \sqrt{{m_n^2-(m_p-m_e)^2\over c_p^2-c_n^2}}
\simeq 
4.1\times 10^{19}\ \left[{10^{-24}\over c_p-c_n}\right]^{1/2}\ \rm eV,}
with $E$ the proton energy in the preferred frame. For this
partcular example, {\it protons with energies exceeding 
$E_2$ are unstable particles that cannot be present among UHE
cosmic rays.}

The above results are expressed in terms of
a nominal choice, $c_p-c_n= 10^{-24}$, lying
beyond the sensitivity of current tests of Lorentz invariance. Thus it is
conceivable that the highest energy cosmic-ray
primaries are stable neutrons.
\medskip

{\it Evading the GZK Cutoff? }
Soon after the discovery of the cosmic background radiation (CBR),
Greisen~\ref\rgr{K. Greisen, Phys. Rev. Lett.  16 (1966) 748.} and
Zatsepin and Kuz'min~\ref\rkz{G.T. Zatsepin and V.A.
Kuz'min, JETP Lett.  41 (1966) 78.}  saw how it limits the
propagation of UHE cosmic rays. Primary nucleons with
sufficient energy will suffer inelastic impacts with CBR photons. 
This  results in what is known as the GZK cutoff, saying that nucleons
with energies $>5\times10^{19}\;$eV cannot reach us from further
than $\sim\!50$~Mpc. However, the primary cosmic-ray
energy spectrum seems to
extend well beyond $10^{20}$~eV~\ref\rcr{AGASA Collab. {\it e.g.,} M. Takeda
{\it et.al.,} Phys. Rev. Lett. 81 (1998) 1163\semi
Fly's Eye Collab. {\it e.g.,} D.J. Bird {\it et al.,} Astrophys. J. 424 (1995
144\semi Haverah Park Collab. See: M.A. Lawrence {\it et al.,} J. Phys. G17
(1991) 733\semi Yakutsk Collab. {\it e.g.,} N.N. Efimov {\it et al.,} in
22nd Intl. Cosmic Ray Conf. (1991) Dublin.}.

The mechanism producing UHE cosmic rays is unknown. Exotic origins have been
suggested, such as: topological defects, active galactic nuclei, and
gamma-ray bursts~ \ref\rother{See \ M. Takeda~\rcr\ and V.  Berezinsky, Nucl.
Phys. Proc.  Suppl. 70 (1999) 419 for further references in these
connections.}. These schemes are constrained or ruled out  by the GZK cutoff.
Other explanations are designed to evade the GZK cutoff: a primary flux of
magnetic monopoles~\ref\rkep{T.W.  Kephart and T.J. Weiler, Astro. Part.
Phys. 4 (1996) 271.}, ``$Z$-boson bursts'' produced by collisions of cosmic
UHE neutrinos off relatively nearby relic neutrinos~\ref\rweiler{T.J. Weiler,
{\tt hep-ph/9710431}.}, and decay products of  hypothetical super-heavy relic
particles~\ref\bere{V. Berezinsky, M. Kachelrie\ss, and V. Vilenkin, Phys.
Rev. Lett.  79 (1997) 4302.}.

We have little to say about the origin of UHE cosmic rays.
Rather, we point out that  there may not be a GZK
cutoff after all. 
Tiny departures from Lorentz invariance, too small to
have been detected otherwise, have effects that
increase rapidly with energy and can 
kinematically prevent cosmic-ray nucleons from
undergoing inelastic collisions with CBR photons.
The cutoff thereby  undone,  a
deeply cosmological origin of UHE cosmic rays becomes 
tenable~\ref\rgzk{A preliminary version of this section appears as S.
Coleman and S.L. Glashow, HUTP-98/A075, {\tt hep-ph/9808446}.}.

To see how the GZK cutoff is affected by 
Lorentz violation, consider the formation reaction yielding
the first pion-nucleon resonance: 
\eqn\eform{ p+ \gamma\ {\rm(CBR)}\longrightarrow \Delta(1232)\,,} 
by which a  nucleon with energy 
$E$ collides inelastically with a CBR photon of energy $\omega$.  
The target photon
energies are 
a thermal distribution with temperature $T=2.73\;$K, or $kT\equiv\omega_0
=2.35\times 10^{-4}\;{\rm eV}$. 
For a head-on impact, the
condition Eq.~\eIIxxi\ determining whether
the reaction can take place  is approximately: 
\eqn\eforma{2\omega +{M_p^2\over 2E}\ge (c_\Delta-c_p)\,E+{M_\Delta^2\over
2E}\,,}
where $c_\Delta-c_p$ is the relevant Lorentz-violating parameter.
If $c_\Delta=c_p$, Eq.~\eforma\  yields the usual threshold
energy for this process,
$E_f=(M_\Delta^2-M_p^2)/4\omega$. Otherwise it 
yields  a quadratic inequality in $E$ which can be satisfied
if and only if $(c_\Delta-c_p)<\hat \delta(\omega)\equiv \omega/2E_f$.
As $c_\Delta-c_p$ is increased toward $\hat \delta$,
the threshold for the formation reaction grows toward $2E_f$. However, if it 
exceeds its critical value,
\eqn\edelcrit{c_\Delta-c_p> 
{2\omega^2\over M_\Delta^2-M_p^2} \simeq 1.7\times
10^{-25}\;[\omega/\omega_0]^2\;,} 
reaction \eform\ becomes  kinematically
forbidden for all $E$.  Recalling that the photon spectrum is thermal, we
see that if $c_\Delta-c_p\sim   \hat{\delta}(\omega_0)$, 
the GZK
cutoff due to resonant $\Delta(1232)$ formation would be relaxed.
Should it much exceed this value, the formation reaction would be precluded 
off virtually all CBR photons.

Reaction \eform\ is the dominant process leading to the GZK
cutoff as originally formulated.
However, if $\Delta(1232)$ formation is not
possible, a
 weakened version of the cutoff
may result from non-resonant photo-production: 
\eqn\eprod{p+ \gamma\ {\rm (CBR)}\longrightarrow p + \pi\,.}
If $c_\pi=c_p$,,
the threshold energy for pion production is
$E_p=M_\pi(2M_p+M_\pi)/4\omega$.  However, as 
$c_\pi-c_p$ is increased from zero,
the  threshold grows. 
As $E\rightarrow \infty$, the pion energy $E_\pi$ 
must remain finite.  Eq.~\eIIxxi\ yields the kinematic condition  for
reaction \eprod\ to occur:
\eqn\kc{ 2\omega \ge  (c_\pi-c_p)\,E_\pi + {m_\pi^2\over 2E_\pi}\,.}
This condition may be satisfied if and only if:
\eqn\eproda{ c_\pi-c_p < \tilde\delta(\omega)\equiv 
{2\omega^2\over m_\pi^2}\simeq
5\times 10^{-24}\ [\omega/\omega_0]^2\,. }
Reaction \eprod, and multiple pion production as well, are
kinematically forbidden {\it at all proton energies\/} if  
$c_\pi-c_p>\tilde\delta(\omega)$.
For the actual case of a thermal gas of photons,
$c_\pi-c_p\sim \tilde \delta(\omega_0)$ would suppress 
photo-pion production, or even eliminate it entirely 
so that no vestige of the cutoff survives.\foot{Note that 
much larger (and experimentally intolerable) violations of
Lorentz invariance would be needed to produce a noticeable effect
on the interactions of UHE
cosmic rays with nuclei in the atmosphere.} 

 At present, lacking detailed observations of the
highest energy cosmic rays and more precise tests of special relativity, we
must regard   as an intriguingly open question  
whether there is a GZK
cutoff, and consequently, whether cosmic rays with energies above the
cutoff can travel cosmological distances.
\medskip

Finally we note that
tiny departures from Lorentz invariance, such as we have
discussed earlier in this Section, 
can  explain the remarkable correlation discovered by Farrar and Biermann:
that the five highest energy cosmic ray events appear to point
toward  
compact radio-loud quasars~\ref\rgf{G.R. Farrar and P.L. Biermann, Phys.
Rev. Lett. 81 (1998) 3579.}.
We suggest that 
these events could have been produced by 
UHE primary {\it neutrons\/} arising from sources at large redshift. 
These particles
could be stable if $c_p>c_n$; they could be immunized
against the GZK cutoff if $c_\pi> c_n$. They are undeflected by
intergalactic magnetic fields because they are neutral.
\bigskip

\newsec{Conclusions}
A wide variety of experiments and observations
offer very precise tests of special relativity. Many of these
results can be interpreted in terms of differences of MAVs
of different particles, such as would result from 
Lorentz-violating (but CPT-conserving) perturbations of the
standard-model Lagrangian. The strongest constraints of this kind 
of which we are aware have been
mentioned earlier and are listed below:
$$\eqalign{ c_p-c_\gamma  <\,& 1\times 10^{-23}\qquad\quad
 \S 1\ \ \quad \rCG\cr 
\vert c_m-c_\gamma\vert   <\,& 6\times 10^{-22}\qquad\quad
 \S 1\ \ \quad \rLam \cr
\big\vert c'-c\big\vert_{\nu_e\nu_\mu}<\,& 6\times 10^{-22}\qquad\quad
  \S 3.2 \quad \bruck\cr
\big\vert c'-c\big\vert_{\mu e}  <\,& 4\times 10^{-21}\qquad \quad
\S 3.1 \quad \rcs\cr
\vert c_{K_L}-c_{K_S} \vert <\,& 3\times 10^{-21}\qquad \quad
 \S 3.3\quad \rsark \cr  }$$
Two of these constraints result from cosmic-ray observations, the others
from experiments performed 
at very high (accelerator) energies,  or in one case, 
at low energy.
They are consistent with strict
Lorentz invariance  to a precision of more than twenty-one  decimal
places.
Possible symmetry violations, 
if present at all, must be exceedingly small. 
How much further must experimenters test special relativity; when is enough
enough? Our analysis addresses this question.

We have seen that maximal velocity differences lying two or three orders of
magnitude  below the current bounds can produce dramatic observable
effects. Some would suppress or forbid  the processes underlying the
GZK cutoff, thereby permitting UHE cosmic-ray nucleons to travel
cosmological distances. Another could stabilize UHE cosmic-ray
neutrons, which would point toward their 
astrophysical sources. Maximal velocity
differences of neutrinos can help to explain the observed properties of
both solar and atmospheric neutrinos. 

Fortunately, much can be done.  Further observations of UHE cosmic
rays  are essential. They may 
confirm a predicted `bump' just below the cutoff~\ref\schr{C.T.
Hill and D.N. Schramm, Phys. Rev.  D31 (1985) 564.}\ resulting from
products of inelastic collisions of primary protons with CBR photons,
thereby providing evidence for the
GZK cutoff. Or,
they could belie the cutoff by  confirming  the Farrar-Biermann
contention~\rgf\  that the highest energy events originate at cosmological
distances. Dedicated searches for velocity oscillations of solar neutrinos,
or of accelerator-produced $\sim\,$TeV neutrinos at baselines of $\sim\!
1000$~km,  could reveal  Lorentz-violating neutrino velocity differences as
small as $10^{-25}$.   Finally, laboratory searches for diurnal
anisotropies more sensitive than any done before have become
feasible~\ref\rfort{E.N. Fortson, private communication.}.

\bigskip\bigskip

\centerline{\bf Acknowledgements}\medskip

We thank Mark Wise for his assistance with  several computations, R. Cowsik
and  B.V. Sreekantan for graciously conveying their results  (as well as
those of V.S. Narasimham and M.R. Krishnaswamy) to us prior to publication, 
Larry Hunter for alerting us to the precision experiment in which he was a
collaborator, and Barry Barish, George Field, Giorgio Giacomelli,
Roman Jackiw, Alan Kosteleck'y, and Jean Zinn-Justin  for
enlightening discussions. This work was supported in part by the
National Science Foundation under grant NSF-PHYS-92-18167.

\bigskip\bigskip
\centerline {\bf Appendix. \ Wrong Reasoning Made Right}
\bigskip
In our first paper on  this subject~\rCG\ we advanced an argument for
neglecting CPT-odd Lorentz-violating interactions.

The argument was based on studies~\ref\rjack{S. Carroll, G. Field and R.
Jackiw, Phys. Rev. D41 (1990) 1231\semi M. Goldhaber and V. Trimble, J.
Astrophys. Astr. 17 (1996) 17. See also: J.F.C. Wardle, R.A. Perley and M.H.
Cohen, Phys. Rev. Lett. 79 (1997) 1801; J.P. Leahy, {\tt astro-ph/9704285};
S. Carroll and G. Field, Phys. Rev. Lett. 79 (1997) 2394.} of an
electromagnetic interaction proportional to
$\half\epsilon_{0ijk}A^iF^{jk}=\vec A \cdot \vec B$. (This term is not
gauge-invariant but it makes a gauge-invariant contribution to the action.)
The new term makes an addition to the space-space part of the photon
self-energy:
$$\Pi_{ij}(p)\rightarrow\Pi_{ij}(p)+i\mu\epsilon_{ijk}p^k,\eqno(A.1) $$
where $\mu$ is a constant with dimensions of mass.  The added term causes
vacuum birefringence; the experimental absence of this effect in
radio-astronomy observations of distant quasars and radio galaxies leads to
a very stringent upper bound, $\mu \leq 10^{-33}\;$eV. In~\rCG\ we argued
that any CPT-odd Lorentz-violating interaction would induce an addition to
the Lagrangian proportional to $\vec A \cdot \vec B$. Thus the extreme
smallness of $\mu$ is evidence for the extreme smallness of all CPT-odd
interactions, and further searches for CPT-violating effects are pointless.
Unfortunately, this argument is invalid.

Let us  consider interactions for which the Lagrange density (not just the
action) is invariant under electromagnetic gauge transformations; for
brevity, we'll refer to these simply as ``gauge-invariant interactions''.
We'll show that to first order in any gauge-invariant CPT-odd interaction,
and to any order in the Lorentz-invariant interactions, $\mu$ vanishes.
(This confirms the expectation that gauge-invariant terms in the
Lagrange density cannot induce gauge-noninvariant ones.)
Thus the fact that $\mu$ is known to be tiny offers no strong constraint on
the magnitude of gauge-invariant CPT-violating interactions.

It's useful to  analyze a simple example before giving our proof in full
generality.  Let $\Gamma_{\mu\nu}(p, q)$ be the one-particle irreducible
(1PI) Green's function for two photons and one Lorentz-violating interaction
Lagrangian. Here the photon with index $\mu$ $(\nu)$ carries momentum $p$
$(q)$.  (Thus the  Lorentz-violating interaction carries momentum $-(p+q)$.)
In first order in the Lorentz-violating interaction
$$\Pi_{\mu\nu}(p)=\Gamma_{\mu\nu}(p,-p).\eqno(A.2)$$
Because the Lorentz-violating interaction is assumed to be
gauge invariant, $\Gamma$ obeys the Ward identities
$$ p^\mu\Gamma_{\mu\nu}(p, q) = 0,  \eqno(A.3a)$$
and
$$ q^\nu \Gamma_{\mu\nu}(p,q) =  0.  \eqno(A.3b)$$

Now let us assume we compute $\Gamma$ by summing only those diagrams that
have no internal photons.  These diagrams have only 
massive-particle\foot{Neutrino  masses may be tiny on the 
scale of high-energy physics, but
they're enormous on the scale of  radio waves.} internal lines; thus, in this
approximation, $\Gamma$ is analytic at $p=q=0$.  Differentiating (A.3a) with
respect to $p^\mu$ and setting $p=0$, we find $\Gamma_{\mu\nu}(0, q)=0$.  Thus
every nonvanishing term in the Taylor expansion of $\Gamma$ has at least one
factor of $p$;  $\Gamma$ is $O(p)$.  The same reasoning applied to (A.3b)
tells us that $\Gamma$ is $O(q)$.  Since $p$ and $q$ are independent
variables, $\Gamma$ is $O(pq)$.  Thus, by (A.2), the addition to the
self-energy  is $O(p^2)$ and makes a vanishing contribution  to $\mu$.

The proof we have given rested on analyticity at vanishing momentum.
Unfortunately, internal photons can produce singularities at precisely this
point.  In the remainder of this appendix we study these singularities and
prove the following:

{\it Theorem:}  Let $\Gamma^{(n)}_{\mu_1\ldots\mu_n}(p_1\ldots p_n)$ denote
the $n$-photon 1PI Green's function, and let us introduce a uniform scaling
parameter $\lambda$ by replacing $p_a$ by $\lambda p_a$.  Then to 
either zeroth or first
order in the CPT-odd interaction, and to any finite order in the
Lorentz-invariant interactions, 
and for any positive $\epsilon$,
$\Gamma^{(n)}$ vanishes as $\lambda$ goes to zero more rapidly than
$\lambda^{n-\epsilon}$. (Note that our announced result, $\mu=0$, is a
corollary of this theorem for $n=2$.)

The  proof proceeds by induction in the number of internal photon lines.

We begin with the case of no internal photon lines.  In this case
$\Gamma^{(n)}$ is analytic at vanishing external momentum, and, to first
order in the Lorentz-violating interactions, the argument is a
straightforward generalization of that for the simple example above. We treat
the $n$ external momenta as independent variables, letting the
Lorentz-violating interaction carry off the momentum inserted at the photon
lines. We then use the Ward identities to show that the  leading term in the
Taylor series is $O(p_1 \ldots p_n) = O(\lambda^n)$.

This doesn't work in zeroth  order in  Lorentz violation, because there is no
Lorentz-violating interaction to carry off the momentum.  The best we can do
is to choose $q_1 \ldots q_{n-1}$ as our independent variables and use the
Ward identities to show that the leading term is $O(p_1\ldots
p_{n-1})=O(\lambda^{n-1})$.  However, the term zeroth  order in Lorentz
violation is Lorentz invariant, and there is no  Lorentz-invariant way to
construct a rank $n$ tensor as a multilinear function of $n-1$ independent
vectors.  Thus this term must vanish, and the  leading term is at best
$O(\lambda^n)$.

To prove the inductive step, we need some information from the once
well-known theory of Feynman-diagram singularities~\ref\elop{R. Eden, P.
Landshoff, D. Olive, and J. Polkinghorne, {\it The Analytic S-Matrix},
Cambridge University Press, 1966.}. For real external momenta, the case of
interest here, the theory can be reduced to a set of simple algorithms:
(1) A given Feynman graph generates a family of reduced graphs, each
obtained by taking some proper subset of the internal lines of the graph and
contracting them to points.  (2) The reduced graph leads to a singularity if
it can be interpreted as a diagram of a classical  process occuring in
space-time, with all particles (that is to say, all uncontracted internal
lines) on the mass shell and moving forward in time.  (3) If there is a
singularity, the associated discontinuity is calculated by the usual Feynman
rules except that $(p^2-m^2+i\epsilon)^{-1}$ is  replaced by
$\delta(p^2-m^2)\theta(p^0)$ in the propagators for the uncontracted internal
lines.

We are now ready to use induction.  We assume the theorem is true for $r$ or
fewer internal photon lines and consider the singularities of graphs with
$r+1$ internal photons.  Since all energies inserted into the graph are
arbitrarily small, there is not enough energy to create a massive particle
and all massive-particle lines must be contracted.  Thus the reduced graph
contains only  photon lines, joined together at vertices which represent
contracted subgraphs of the original graph.  If we sum, in some fixed order
of perturbation theory, all graphs leading to the same reduced graph, the
vertices become the $\Gamma$'s at the appropriate order of pertubation
theory.  (We're being a bit sloppy here:  subgraphs of a 1PI graph need not
be 1PI, so the vertices can contain tree graphs.  However, it's easy to check
that these have  no effect on the power-counting of the next paragraph.)

Because some of the internal photons in the original graph must be
uncontracted, an $m$-line vertex in the reduced graph must have $r$ or fewer
internal photon lines and,  by the inductive hypothesis, must  vanish at small
$\lambda$ more rapidly than $\lambda^{m-\epsilon}$. It will be convenient to
consider this as one factor of $\lambda^{1-\epsilon}$ for each photon  line
attached to the vertex.  We can now compute the $\lambda$-dependence of the
discontinuity. Every external photon contributes a factor of
$\lambda^{1-\epsilon}$.  Every internal photon contributes a factor of
$\lambda^{2-\epsilon}$ from its two ends and a factor of $\lambda^{-2}$
from the $\delta$-function, yielding no  net contribution.  Every independent
loop integration contributes a factor of $\lambda^4$.  Thus the discontinuity
vanishes faster than $\lambda^{n+4L-\epsilon}$, where L is the number of
loops.  This in turn vanishes faster than $\lambda^{n-\epsilon}$, which  is
the result we need.  (It is critical that we are computing
the discontinuity and not the full $\Gamma$.  The $\delta(p^2)\theta(p^0)$
propagators in the reduced graph keep the internal momenta small when the
external momenta are small and legitimize the use of 
a small-momentum bound for
the vertices.)

Once we have the discontinuity, we can construct a function with that
discontinuity, for example, by integrating a dispersion relation.  This
function also vanishes faster than $\lambda^{n-\epsilon}$.
Thus $\Gamma^{(n)}$ is the sum of a singular function that vanishes faster
than $\lambda^{n-\epsilon}$ and a function that is free of singularities,
that is to say, an analytic function.  But we can
use the same arguments 
for the analytic function that we used 
for the case of no internal photon lines to
show that it vanishes like $\lambda^n$.  This  completes the proof.

\listrefs
\bye